\documentclass[a4paper]{jpconf}
\usepackage{graphicx}
\begin{document}
\title{Experimental  determination  of  gravitomagnetic  effects  by  means  of  ring  lasers}

\author{Angelo Tartaglia}

\address{Politecnico, corso Duca degli Abruzzi 24, 10129 Torino, Italy, and INFN}

\ead{angelo.tartaglia@polito.it}

\begin{abstract}
A new experiment aimed to the detection of the gravito-magnetic Lense-Thirring effect at the surface of the Earth will be presented; the name of the experiment is GINGER. The proposed technique is based on the behavior of light beams in ring lasers, also known as gyrolasers. A three-dimensional array of  ringlasers will be attached to a rigid "monument"; each ring will have a different
orientation in space. Within the space-time of a rotating mass the propagation of light is indeed anisotropic; part of the anisotropy is purely kinematical (Sagnac effect), part is due to the interaction between the gravito-electric field of the source and the kinematical motion of the observer (de Sitter effect), finally there is a contribution from the gravito-magnetic component of the Earth (gravito-magnetic frame dragging or Lense-Thirring effect). In a ring laser a light beam traveling counterclockwise is superposed to another beam traveling in the opposite sense. The anisotropy in the propagation leads to standing waves with slightly different frequencies in the two directions; the final effect is a beat frequency proportional to the size of the instrument and its effective rotation rate in space, including the gravito-magnetic drag. Current laser techniques and the performances of the best existing ring lasers allow at the moment a sensitivity within one order of magnitude of the required accuracy for the detection of gravito-magnetic effects, so that the objective of GINGER is in the range of feasibility and aims to improve the sensitivity of a couple of orders of magnitude with respect to present. The experiment will be underground, probably in the Gran Sasso National Laboratories in Italy, and is based on an international collaboration among four Italian groups, the Technische Universit\"{a}t M\"{u}nchen and the University of Canterbury in Christchurch (NZ).
\end{abstract}

\section{Introduction}

The theory of relativity, both special (SR) and general (GR), tells us that the propagation of light along closed space contours appears to be anisotropic either when the observer is rotating or when the gravitational field belongs to a rotating mass, or of course when both circumstances are present. The purely kinematical effect had initially also a classical description and is known as the Sagnac effect \cite{sagnac}; it is easy to measure by modern interferometric techniques or using ring-lasers. The other two effects, due to the gravitational field of a rotating mass, arise from the coupling of the gravito-electric part of the field with the rotation of the observer (geodetic or de Sitter or Schiff effect, according to the experimental circumstances) \cite{desitter} and from the gravito-magnetic component of the field (Lense-Thirring effect) \cite{thirring}. I shall call the two latter phenomena physical or GR effects.

The physical rotation effects have been measured so far by two experiments in space. In both cases the precession induced by the rotation of the earth on the axis of a gyroscope has been used. By this means the Gravity Probe B experiment, stemmed from an idea proposed by Schiff \cite{schiff}, verified the terrestrial geodetic effect with a 0.28\% accuracy and the Lense-Thirring drag with a 19\% accuracy \cite{gpb}. Another measurement has been made by Ciufolini using the data from the laser ranging of the LAGEOS satellites, launched for a different purpose. From the precession of the orbits of the LAGEOS the Lense-Thirring effect has been confirmed at 10\% accuracy \cite{lageos}. Another experiment based on the laser ranging technique is presently under way: the LARES satellite has been launched on February 13th 2012 on a dedicated mission and is now collecting data with the purpose of measuring the Lense-Thirring at a 1\% accuracy \cite{lares}. These experiments are not easy and have the inconveniences of being in space.

What I am presenting here is a different opportunity to measure the GR effects of rotation resting on the surface of the planet earth (actually under the surface, in an underground location). The proposal is to use light as a probe of the configuration and properties of space-time. The device to be used is a ring-laser (actually a three-dimensional array of ring-lasers). The name of the experiment is GINGER and it will be described in the following. The GINGER proposal is the result of a collaboration whose Principal Investigator is Angela Di Virgilio of the Pisa section of the Italian INFN and involving primarily groups from a number of Italian institutions (the universities of Florence, Naples, Padua, Pisa, Turin-Politecnico; various INFN sections) and from the German Technische Universit\"{a}t M\"{u}nchen. A consultancy link is also active with a group of the Canterbury University of Christchurch (NZ). The detailed proposal is presented in \cite{ginger}.

\section{Propagation of light in the field of a rotating mass}

The line element of the space-time of a steadily rotating mass may in general be written as:

\begin{equation}
\label{linea}
ds^2=g_{00}dt^2+g_{rr}dr^2+g_{\theta\theta}d\theta^2+g_{\phi\phi}d\phi^2+2g_{0\phi}dtd\phi
\end{equation}
where polar coordinates in space are used, to evidence the symmetry. The $g_{\mu\nu}$'s are the elements of the metric tensor and each of them does not depend on $t$ and $\phi$ due to the symmetry. The mixed term $g_{0\phi}$ accounts for the rotation of the source, i.e. the central mass, and is responsible for the so called gravito-magnetic effects; $g_{0\phi}$ cannot be eliminated by any global coordinate transformation.
Had we used different coordinates, we would have had in general three elements  $g_{0i}$, with the index $i$ labeling the three space coordinates. The three $g_{0i}$'s may be read as the components of a three-vector $\overrightarrow{h}$, which can be interpreted as the vector-potential of a gravito-magnetic field $\overrightarrow{B}_g$, i. e. of the rotation depending part of the gravitational field \cite{gmag}. In conventional three-dimensional notation it is:

\begin{equation}
\label{rot}
\overrightarrow{B}_g=\overrightarrow{\nabla}\wedge\overrightarrow{h}
\end{equation}

Such a gravito-magnetic field has the configuration of a dipole. Let us then evaluate the time of flight (TOF) of a light beam constrained by some physical system to travel along a closed space path; the TOF is the proper time interval of a laboratory within which the experiment is performed. The result is obtained from eq. (\ref{linea}) considering a null spacely closed world-line and integrating along it. The only term sensitive to the rotation sense is the one containing $g_{0\phi}$ that linearly multiplies $d\phi$, which is odd in the angle; that term will be the only one left when subtracting the clockwise from the counter-clockwise travel time. In formulae it is:

\begin{equation}
\label{tof}
\delta T=T_+-T_-=-2\sqrt{g_{00}}\oint\frac{g_{0\phi}}{g_{00}}d\phi
\end{equation}

\section{Ring lasers}

\begin{figure}[h]
\begin{center}
\includegraphics[height = 80 mm, width=120 mm]{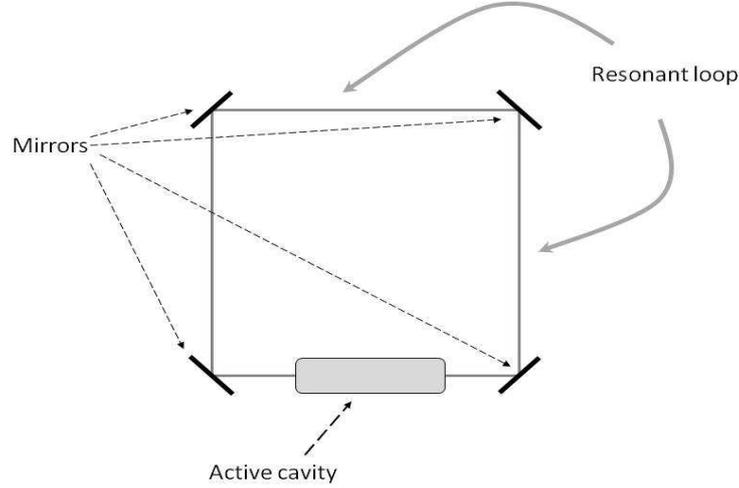}
\caption{\label{fig:loop}The schematic view of a square ring laser is shown. The right- and left-handed beams exiting the active cavity are led to form a square resonant loop by four mirrors.}
\end{center}
\end{figure}

The anisotropy pointed out in the previous section is the basis of the operation of a ring laser. Consider the scheme shown on fig. \ref{fig:loop}. An active cavity produces two light beams traveling in opposite directions; four mirrors deviate the beams to form a closed square path in space. In order to have a ring the mirrors cannot be less than three but of course could be more.

The loop forms in turn a resonant cavity. According to formula (\ref{tof}) there is a time of flight difference between the two beams. Since we suppose to be in a stationary condition we have a pair of standing waves; each of them is formed by an integer number of wavelengths, but, due to the TOF difference (\ref{tof}), the clockwise and counter-clockwise wavelengths are slightly different from one another. In the lowest mode the integer is the same for both beams and the superposition leads to a beat note whose frequency is proportional to the TOF difference \cite{ginger}:

\begin{equation}
\label{beat}
f_b=\frac{c^2}{\lambda P}\delta T
\end{equation}
$P$ is the length of the loop and $\lambda$ is the fiducial wavelength  of the laser. The beat frequency can be extracted and read at one of the corners of the ring.

In the case of an earth-bound laboratory the relevant metric tensor elements can be approximated to the lowest significant order including the angular momentum of the earth, obtaining:

\begin{eqnarray}
g_{0\phi}&\simeq&(2\frac{GJ_\oplus}{c^3r}-r^2\frac{\omega}{c}-2G\frac{M_\oplus}{c^2}\frac{r\Omega_\oplus}{c})sin^2{\theta} \nonumber \\ \label{approx} \\
g_{00}&\simeq&1-2\frac{GM_\oplus}{c^2r}-\frac{\omega^2r^2}{c^2}sin^2{\theta} \nonumber
\end{eqnarray}

In (\ref{approx}) the label $_\oplus$ designates physical quantities belonging to the planet Earth: the total mass $M$, the angular momentum $J$ and the angular velocity $\Omega$; $\omega$ is the angular velocity of the apparatus and in the following we shall assume that $\omega = \Omega_\oplus$; $\theta$ is the colatitude of the laboratory and $r$ is the distance from the center of the earth.

Using the above approximation it is possible to write down the expected signal in the form of the beat frequency \cite{ginger}:

\begin{equation}
\label{signal}
\delta f = 4\frac{A}{\lambda P}[\overrightarrow{\Omega}_\oplus-2\frac{GM_\oplus}{c^2R}\Omega_\oplus sin{\theta} \hat{u}_\theta + \frac{GJ_\oplus}{c^2R^3}(2\cos{\theta} \hat{u}_r+ \sin{\theta} \hat{u}_\theta)]\cdot \hat{u}_n
\end{equation}
$A$ is now the area contoured by the light beams; $R$ is the radius of the earth; the unit vectors $\hat{u}_r$, $\hat{u}_\theta$ and $\hat{u}_n$ are respectively: radial, along the local meridian in the sense of increasing colatitude, perpendicular to the plane of the ring (provided it is in a plane). The whole quantity in front of the square bracket is called the scale factor: the bigger it is the stronger is the signal from the apparatus.

Within the square brackets of formula (\ref{signal}) the three terms represent three real or effective angular velocities. The first is the rotation rate of the earth and accounts for the kinematical Sagnac effect; the second term corresponds to the geodetic or de Sitter effect; the third and last term is the gravito-magnetic contribution and accounts for the Lense-Thirring effect. The two physical terms (as they are also called) in (\ref{signal}) are of the same order of magnitude and nine orders of magnitude smaller than the dominant Sagnac term.

\subsection{Commercial gyrolasers}

\begin{figure}[h]
\includegraphics[width=14pc]{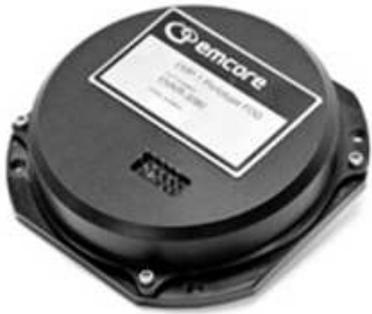}\hspace{2pc}%
\begin{minipage}[b]{14pc}\caption{\label{fig:gyrol}A commercial optical fiber gyrolaser.}
\end{minipage}
\end{figure}

Ring lasers are already in use as "gyrolasers" for the measurement of rotation rates, for instance in airplanes or in submarines. They are replacing the mechanical gyroscopes formerly used for the same purpose, whence the name of gyrolasers. Commercial gyrolasers are compact objects with sensitivities in the order of $\sim 10^{-7} rad/s/\surd Hertz$ obtained with appropriate values of the scale factor resulting from the use of multiply wound optical fibers. Often one single device is made of three gyrolasers aligned along three mutually perpendicular axes in order to sense the full angular velocity three-vector. Fig. (\ref{fig:gyrol}) shows an example of commercial gyrolaser.

\subsection{G-Pisa}

\begin{figure}[h]
\begin{center}
\begin{minipage}{15pc}
\includegraphics[width=15pc]{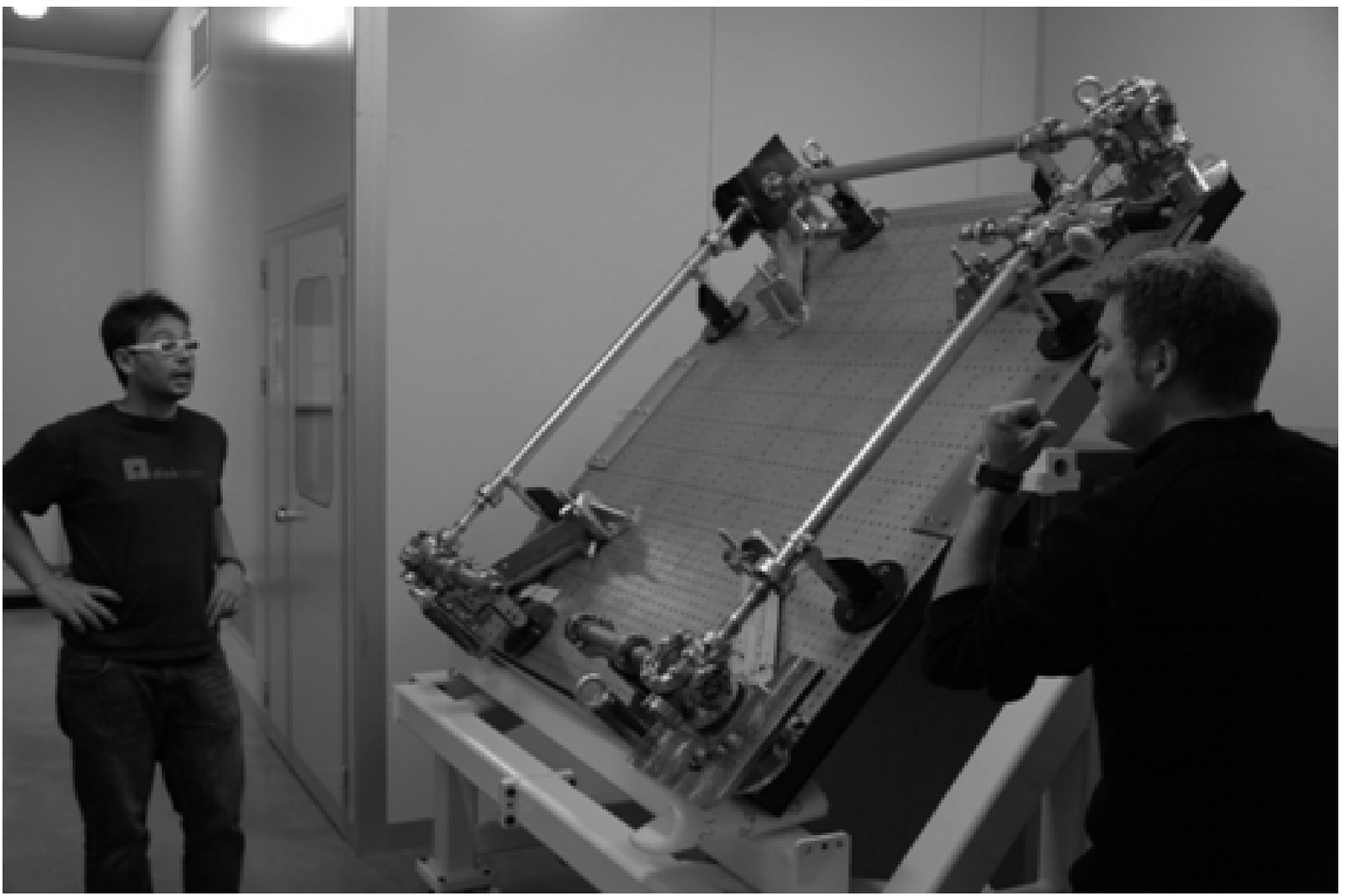}
\caption{\label{fig:GPisa}Picture of the G-Pisa ring laser on its granite support in a tilted configuration.}
\end{minipage}\hspace{3pc}%
\begin{minipage}{15pc}
\includegraphics[width=15pc]{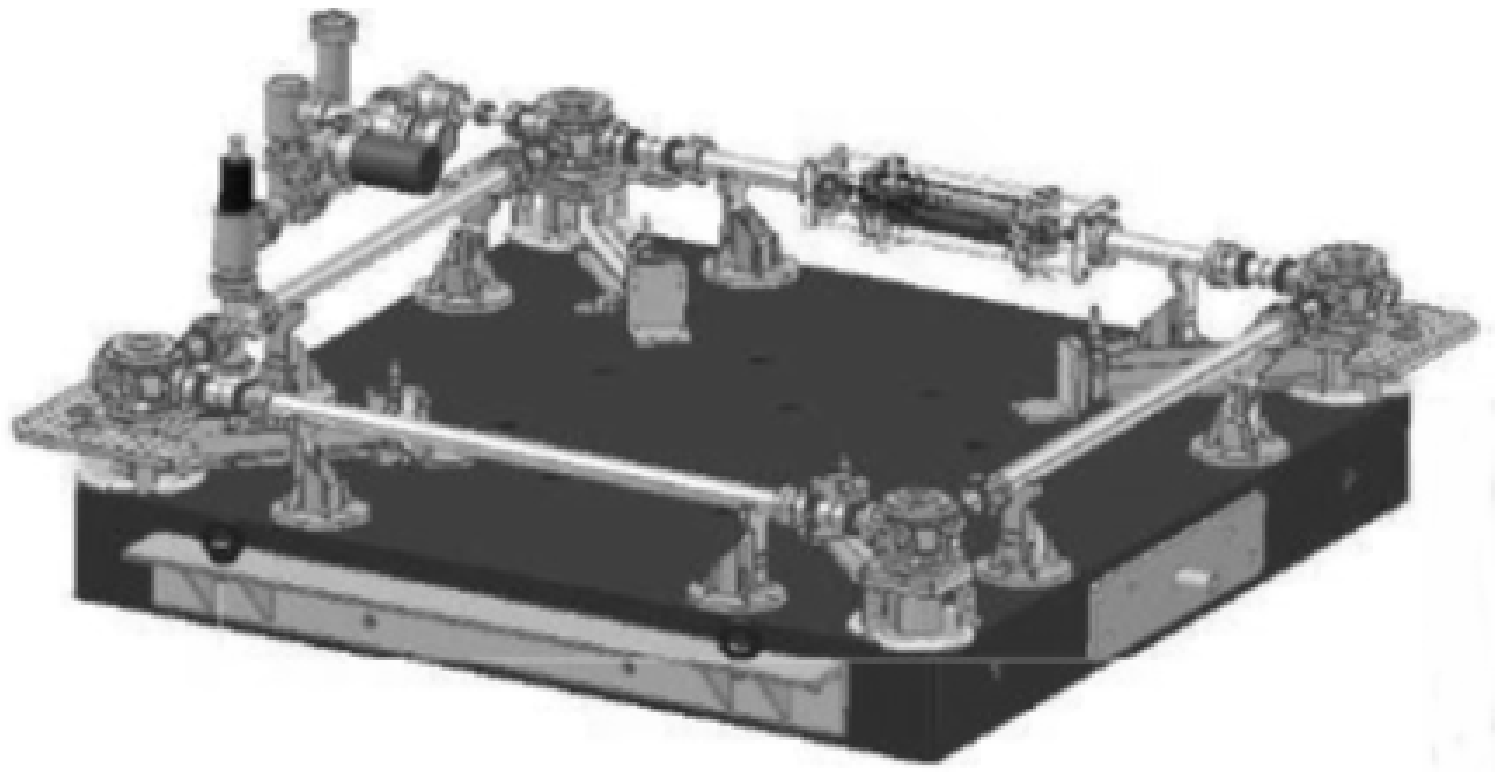}
\caption{\label{fig:Schema}Schematic view of G-Pisa. The light beams are contained into tubes filled of low pressure He gas; the mirrors are mounted at the four corners of the device; the active laser cavity is visible; the extraction of the signal happens at the left upper corner.}
\end{minipage}
\end{center}
\end{figure}

The accuracy of commercial gyrolasers is not enough to perform scientific experiments and can hardly detect the diurnal rotation of the earth. For scientific purposes more refined and bigger instruments are required. An example is G-Pisa, initially developed to test the rotational stability of the Virgo gravitational interferometer located at Cascina near Pisa in Italy. G-Pisa is a square ring with a 1.35 m long side; it is visible in fig.s \ref{fig:GPisa} and \ref{fig:Schema}, both in its real aspect and in a schematic drawing.

The source of light is a He-Ne laser with an adjustable output power of a few tens of nW. The device is mounted on a thick granite table that can be held at various pitches from horizontal to vertical.

The sensitivity of G-Pisa  is somewhat in between $10^{-9}$ and $10^{-10}$ $rad/s/\surd Hertz$ \cite{pisa}. It is not enough to sense the GR effects, but permits to measure, besides the rotation of the earth, many interesting motions of the surface of the earth at the laboratory, of geophysical origin. G-Pisa is destined to become a test-bed for technologies in preparation for the future GINGER instrument.

\subsection{The Cashmere cavern instruments}

\begin{figure}[h]
\begin{center}
\includegraphics[width=18pc]{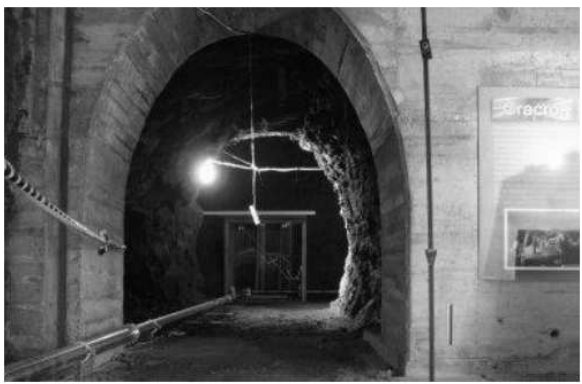}\hspace{2pc}%
\begin{minipage}[b]{14pc}\caption{\label{fig:NZ}View of one of the vacuum pipes hosting the laser beam of a ring laser in the Cashmere cavern near Christchurch, NZ.}
\end{minipage}
\end{center}
\end{figure}

During the years a number of ring lasers for fundamental research have been built by a group of the University of Canterbury in Christchurch, NZ. The first one was C-I \cite{stedman}, others followed up to UG-2 whose interclosed area is 834 $m^2$ \cite{ug2}. They were all located underground in the Cashmere cavern, near Christchurch (see fig. \ref{fig:NZ}).

Unfortunately the important gain in the scale factor obtained thanks to the size of the apparatus is overridden by the mechanical instabilities due to the length of the arms, so UG-2 did not permit measurements of the GR effects.

The laboratory in the Cashmere cavern suffered relevant damages from the severe earthquakes that shook the Christchurch area in September 2010 and February 2011, so that now it is not in operation. For the future however the group in the University of Canterbury is planning to build a new triangular ring laser with 5 m sides, mounted on a rigid support, that promises very good performances.

\subsection{G in Wettzell}

\begin{figure}[h]
\begin{center}
\begin{minipage}{15pc}
\includegraphics[width=15pc]{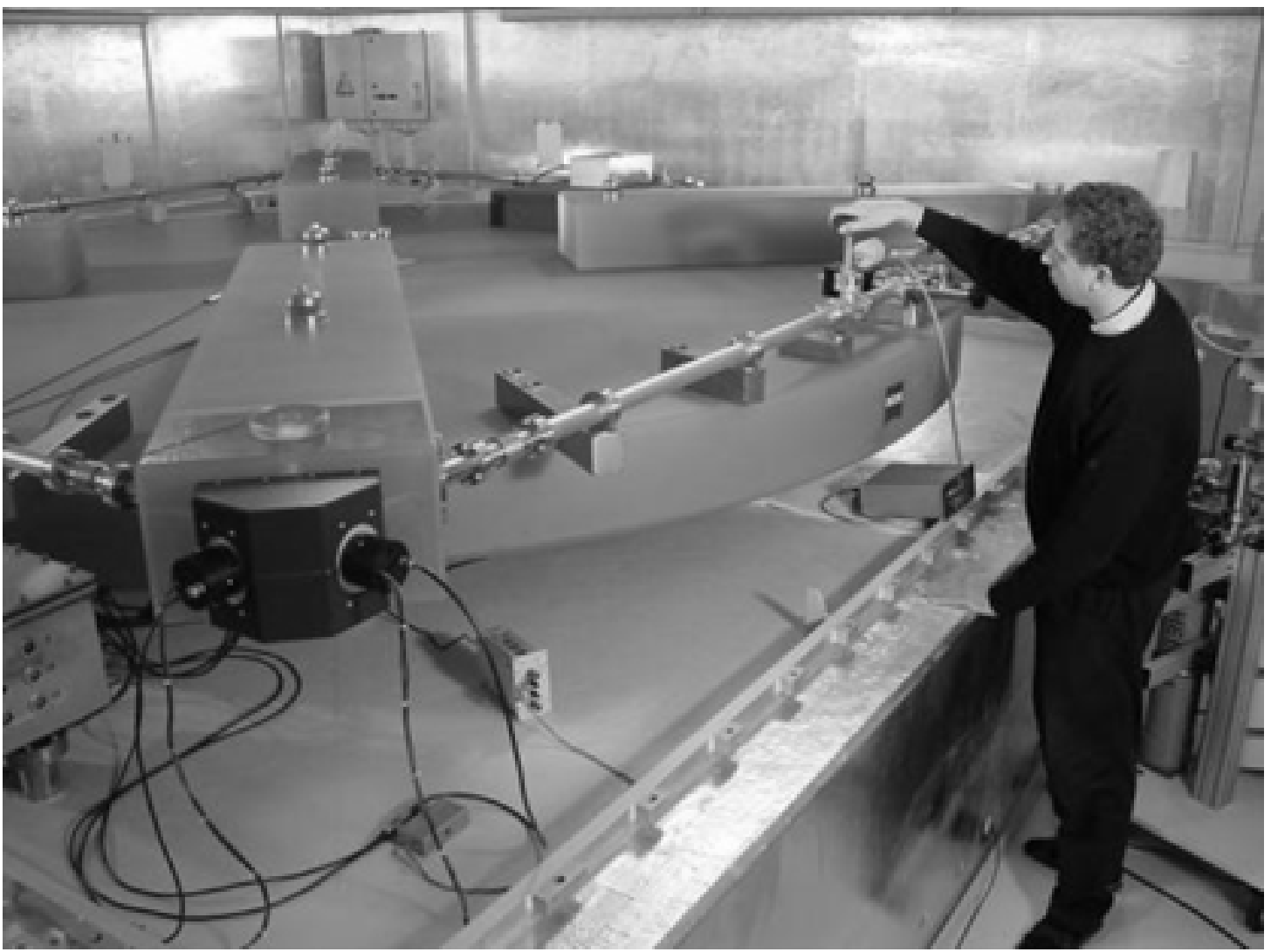}
\caption{\label{fig:wettzell}Picture of the G ring laser located in Wettzell on its zerodur support; the apparatus is shown without the temperature and pressure controlled casing that covers it when in operation.}
\end{minipage}\hspace{3pc}%
\begin{minipage}{15pc}
\includegraphics[width=15pc]{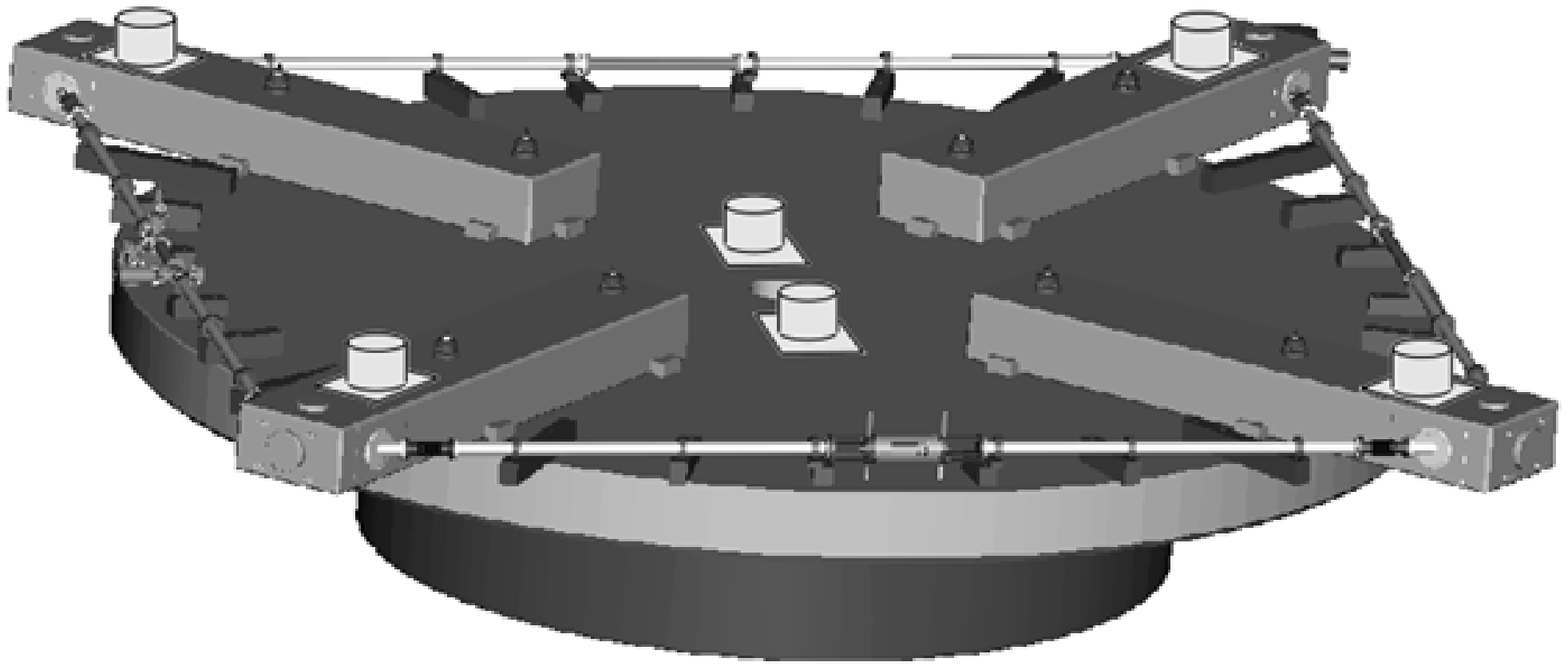}
\caption{\label{fig:G}Schematic view of the G ring laser. The device is mounted on a single block of zerodur, an extremely rigid and thermally stable ceramic material}
\end{minipage}
\end{center}
\end{figure}

The best ring laser in the world is at the moment the G ring at the Geod\"{a}tisches Observatorium Wettzell in Bavaria. It is a square ring, 4 m side, mounted on a monolithic block made of zerodur, a rigid ceramic material with a very small thermal expansion coefficient.The source of light is a He-Ne laser with 20 nW output power. The laboratory is under an artificial mound, in order to isolate it as far as possible from environmental disturbances; the apparatus is in turn kept within a chamber where pressure and temperature are stabilized; the zerodur table rests on a concrete pillar attached to the natural rock pavement under ground. Fig.s \ref{fig:wettzell} and \ref{fig:G} present a picture and a scheme of the instrument.

\begin{figure}[h]
\begin{center}
\includegraphics[height = 70 mm, width=120 mm]{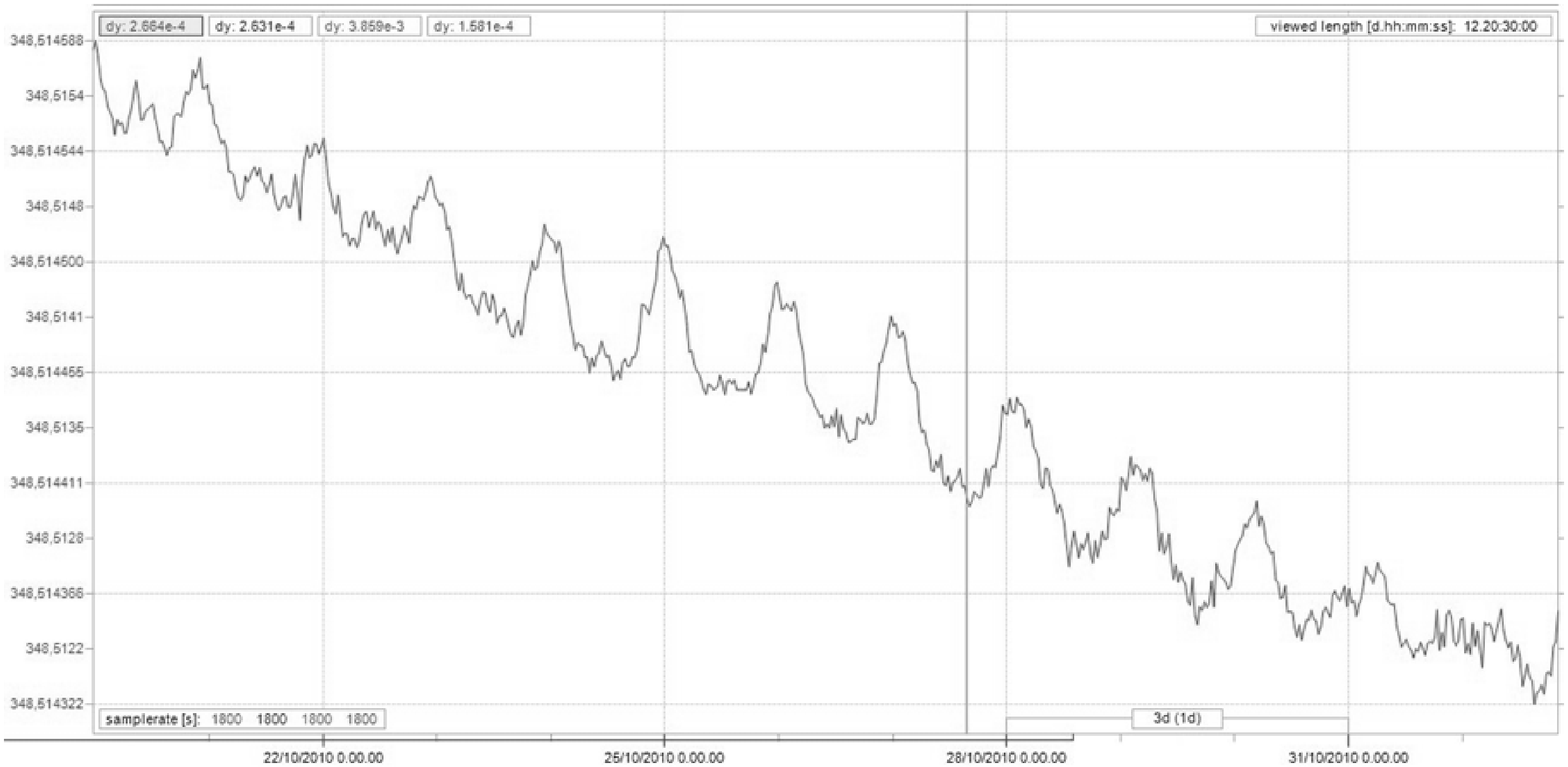}
\caption{\label{fig:grafo}Signal from G during 13 days. Superposed to the Sagnac signal it can easily be seen the diurnal polar motion of the rotation axis of the earth; the height of individual peaks is a bit less than 50 $\mu Hz$.}
\end{center}
\end{figure}

The sensitivity of G arrives to $4.5\times 10^{-12}$ $rad/s/\surd{Hz}$ and is already close to the value required in order to reveal the GR effects of rotation. Fig. \ref{fig:grafo} shows an example of data taken by G during thirteen days. The peaks correspond to the diurnal polar motion of the terrestrial axis; each of them has a height of the order of 50 $\mu Hz$ or less, i.e. $\sim 10^{-8}$ times the Sagnac signal above which the peaks rise.

The accuracy of G is within one order of magnitude from the expected physical rotation terms. The instrument is an excellent sensor of any sort of geophysical rotations as well as of tiny pushes from environmental interactions \cite{Gw}.

\section{GINGER}

As we have seen, G in Wettzell is already close to the sensitivity needed for gravito-magnetic measurements so that an additional effort allowed by present laser technology can make an earth experiment to reveal the Lense-Thirring effect feasible. To this purpose the collaboration presented in the introduction is proposing a new experiment relying on the expertise of all members: GINGER.

GINGER will not be a single ring laser, but rather a three-dimensional array of rings. Each ring will be a square loop with a 6 m long side. The minimum number of rings is three, with the three normals oriented along mutually perpendicular directions in space; in this way the three space components of the effective angular velocities would simultaneously be measured.

\begin{figure}[h]
\begin{center}
\includegraphics[height = 65 mm, width=120 mm]{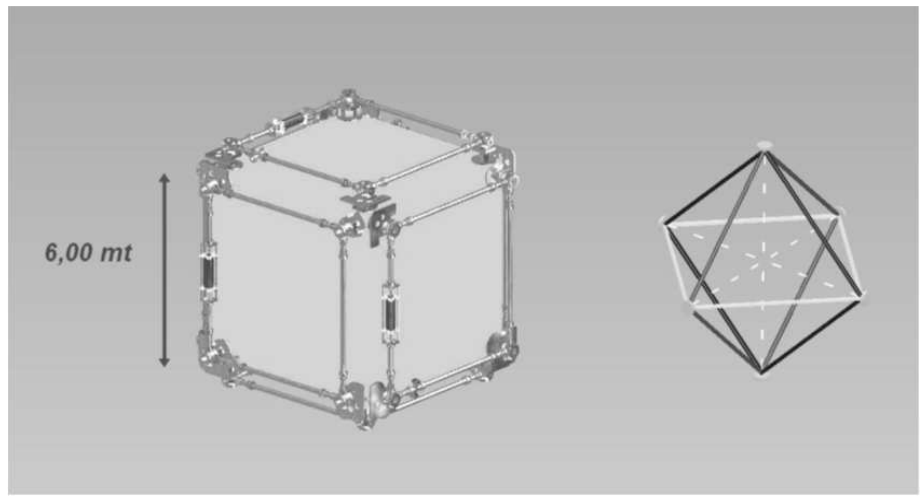}
\caption{\label{fig:gingero}Two possible configurations for GINGER. On the left a cubic concrete monument carries six ring lasers, perpendicular to each other in pairs; on the right an octahedral structure is presented with three mutually perpendicular square rings.}
\end{center}
\end{figure}

A possibility is to actually have six loops, coupled in pairs of equal orientation; the consequent redundancy would allow a better control of noise. Fig. \ref{fig:gingero} shows two possible configurations of GINGER.

One is a cube whose faces support pairs of equal square loops; the other is an octahedron, carrying three square loops. The octahedral structure allows a better control of the geometry through the diagonals, along which three Fabry-P\'{e}rot interferometers would be placed.
In all cases the mirrors will be attached to a concrete "monument" and the stability of the configuration will be insured more through active control of the position of the mirrors than via passive rigidity of the support.

The design power of the light source is 200 nW and the quality factor of the cavity is $Q = 3\times 10^{12}$.

The site for GINGER could probably be the underground Gran Sasso national laboratory (LNGS) of the INFN in Italy where the appropriate technological facilities are already available and more than 1000 m of rock above the apparatus will be a most effective screen against surface noise.

The purpose is to measure the Lense-Thirring effect with a 1\% accuracy after one year of integration.

\section{Conclusion}

We have seen how ring lasers can be used for fundamental tests of general relativity on earth, having attained, in recent years, an unprecedented accuracy. A synergy between three groups working on ring lasers in the world has brought to the proposal of a dedicated new experiment named GINGER, whose technological improvements will indeed allow for the measurement of the general relativistic effects originated from the rotation of the mass of the earth. Actually if the accuracy will be good enough it is also possible to arrive to put constraints on a couple of PPN parameters. In fact the PPN form of three physical effective rotations that GINGER could reveal is:

\begin{eqnarray}
\overrightarrow{\Omega}_G & = & -(1+\gamma)\Omega_\oplus \frac{GM}{c^2R}\sin{\theta}\hat{u}_\theta \nonumber \\
\overrightarrow{\Omega}_B & = & -\frac{1+\gamma+\alpha_1 /4}{2} \frac{G}{c^2R^3}(\overrightarrow{J}_\oplus - 3(\overrightarrow{J}_\oplus \cdot \hat{u}_r)\hat{u}_r) \label{PPN} \\
\overrightarrow{\Omega}_W & = & - \frac{\alpha_1}{4}\frac{GM}{c^2R^2}\hat{u}_r\wedge \overrightarrow{W} \nonumber
\end{eqnarray}

The $\gamma$ and $\alpha_1$ parameters would account for possible anomalous (from the view point of GR) dependencies of the curvature from the mass of the source and for possible preferred frame effects; the $\overrightarrow{W}$ appearing in formula (\ref{PPN}) is a gravitational three-vector potential associated with the presence of a preferred frame. In GR it is $\gamma=1$ and $\alpha_1=0$.

GINGER has got a principle approbation from the Italian INFN. The seismic noise of the possible location in the LNGS has been checked at the beginning of 2011 by a German team; the G-Pisa ring will be moved, at the beginning of 2013, to the LNGS in order to characterize the place from the viewpoint of the rotational noise. Meanwhile a new ring approximately the same size as G-pisa will be built in order to serve as a test bed for mirrors stability and control.

Summing up GINGER is on a track that will lead it to completion and we are confident that its performances will match up with the promises and expectations, paving the way for a number of future investigations using light as an intrinsically relativistic probe for the structure of space-time.

\ack{This presentation has been given on behalf of the whole GINGER collaboration mentioned in the introduction and in particular of the authors of ref. \cite{ginger}.}

\end{document}